\documentclass[prl,twocolumn,superscriptaddress,showpacs,amsmath,amssymb]{revtex4-1}
\usepackage{graphicx}
\usepackage{amsmath}
\usepackage{amssymb}
\usepackage{float}

\usepackage[dvipsnames]{xcolor}

\usepackage{cancel,xcolor}
\usepackage{indentfirst}
\usepackage{CJKulem}
\usepackage{soul,xcolor}
\setstcolor{red}
\usepackage{ulem}
\usepackage{cancel}
\usepackage[pdfpagemode=UseNone,pdfstartview=FitH,colorlinks=true,linkcolor=blue,urlcolor=blue,anchorcolor=blue,citecolor=blue]{hyperref}

\newcommand{\G}{{\mbox{\boldmath{${\cal G}$}}}}

\def\sig{{\mbox{\boldmath{$\sigma$}}}}
\def\mubold{{\mbox{\boldmath{$\mu$}}}}

\def\Aa2#1{\textcolor{magenta}{#1}}

\def\Aa1#1{\textcolor{blue}{#1}}

\def\prb#1{{ Phys.\ Rev. B\/} {\bf#1}}

\def\prl#1{{ Phys.\ Rev.\ Lett.} {\bf#1}}

\begin{document}

\title{Magnetoconductance Anisotropies and Aharonov-Casher Phases }


\author{R. I. Shekhter}
\affiliation{Department of Physics, University of Gothenburg, SE-412
96 G{\" o}teborg, Sweden}

\author{O. Entin-Wohlman}
\email{orawohlman@gmail.com}
\affiliation{School of Physics and Astronomy, Tel Aviv University, Tel Aviv 69978, Israel}

\author{M. Jonson}
\affiliation{Department of Physics, University of Gothenburg, SE-412
96 G{\" o}teborg, Sweden}

\author{A. Aharony}
\affiliation{School of Physics and Astronomy, Tel Aviv University, Tel Aviv 69978, Israel}


\begin{abstract}
%
The spin-orbit interaction (SOI) is a key tool  for manipulating and functionalizing  spin-dependent electron transport.
The desired function often depends on the SOI-generated phase that is accumulated by the wave function of an electron as it passes through the device. This
phase, known as the Aharonov-Casher phase, therefore depends on both the device geometry and the SOI strength. Here we propose a method for directly measuring the  Aharonov-Casher phase generated in an SOI-active weak link, based on the Aharonov-Casher-phase dependent anisotropy of its magnetoconductance. Specifically we consider weak links in which
the Rashba interaction is caused by  an external electric field, but our method is expected to apply also for other forms of the spin-orbit coupling.
Measuring this magnetoconductance  anisotropy thus allows calibrating   Rashba spintronic devices by an external electric field that tunes the spin-orbit
interaction and hence the Aharonov-Casher phase.

\end{abstract}


\date{\today}
\maketitle

\noindent{\bf Introduction.} The spin-orbit interaction (SOI), which allows for an interplay between charge and spin currents in
all-electrical devices \cite{Sahoo,Hirohata}, substantially  affects the  transport properties of two- and three-dimensional  conductors.
The spin-Hall effect \cite{Wunderlich}, the
electrical control of the magnetization in magnetic heterostructures
by interfacial spin-orbit toques \cite{Amin}, and the spin relaxation  yielding quantum anti weak-localization  in impure conductors \cite{Bergmann,Iizasa} are  examples of such an influence.

The SOI is due to the interaction between the magnetic moment  $\mubold$ of a particle moving with velocity ${\bf v}$ through a static electric field {\bf E} and the magnetic field {\bf B}$_{\rm so}=(\bf{E}\times\bf{v})/$c$^2$ seen in the rest frame of the particle. This relativistic effect adds a term
%
\begin{equation}
\label{E1}
{\cal H}_{\rm so}(k) = \mubold \cdot {\bf B}_{\rm so} = \mubold \cdot ( {\bf E} \times {\bf v})/c^2.
\end{equation}
to the Hamiltonian, which can be removed by a gauge transformation, $\Psi \to \exp(i\hat\phi_{\rm AC})\Psi$,
that adds a phase factor  to the
wave function. Here
\begin{equation}
\label{E2}
\hat\phi_{\rm AC} = -\frac{1}{\hbar} \int^{t} \mubold \cdot {\bf B}_{\rm so} dt' = \frac{1}{\hbar c^2} \int^{\bf r} ( {\bf E} \times {\mubold})\cdot d{\bf r}' ,
\end{equation}
where the integration is over the path of the particle, is known as the Aharonov-Casher phase  \cite{AC,Qian}.
We use the notation $\hat\phi_{\rm AC}$ for the phase operator and $\phi_{\rm AC}$ for its eigenvalue, to be discussed below.

One notes from Eqs. (\ref{E1}) and (\ref{E2}) that the strength of the SOI, which can be readily measured \cite{Nitta,Nowack},
is a local property while, importantly, the Aharonov-Casher phase depends in addition on the geometry of
the device, usually not precisely known, and is therefore a global property of the device.

%
The Aharonov-Casher phase gives rise to quite a number of new spintronic functionalities in devices  formed by weak links bridging bulk conductors in configurations  where the spin-orbit interaction is active solely in the weak link \cite{RIS2017}.
Such devices are capable of 
spin-selective electric transport \cite{Varela,Comment,Utsumi}. They allow for
spin accumulation and generation of spin currents  \cite{Takahashi,Davidson,Radic} and for
electromotive forces  \cite{photo} and spin-polarization of superconducting Cooper pairs \cite{RIS16}.
These  effects are 
more striking in quantum networks built of spin-orbit active  weak links
\cite{Konig,Frustaglia} where  the Aharonov-Casher phase  dominates
the interference of the electronic spinors \cite{Rudriguez} and even yield  spin filtering \cite{Matityahu}.

Because of its importance it would be useful to be able to measure and tune the Aharonov-Casher phase
without separate knowledge of the SOI strength and device geometry.
 The purpose of this Letter is to propose a method for achieving just that.
Our proposal is based on magnetoconductance measurements and does not require observing interference patterns
in multiply connected systems.
%


\begin{figure}
\centering
\vspace{-1cm}
\includegraphics[width=.5\textwidth]{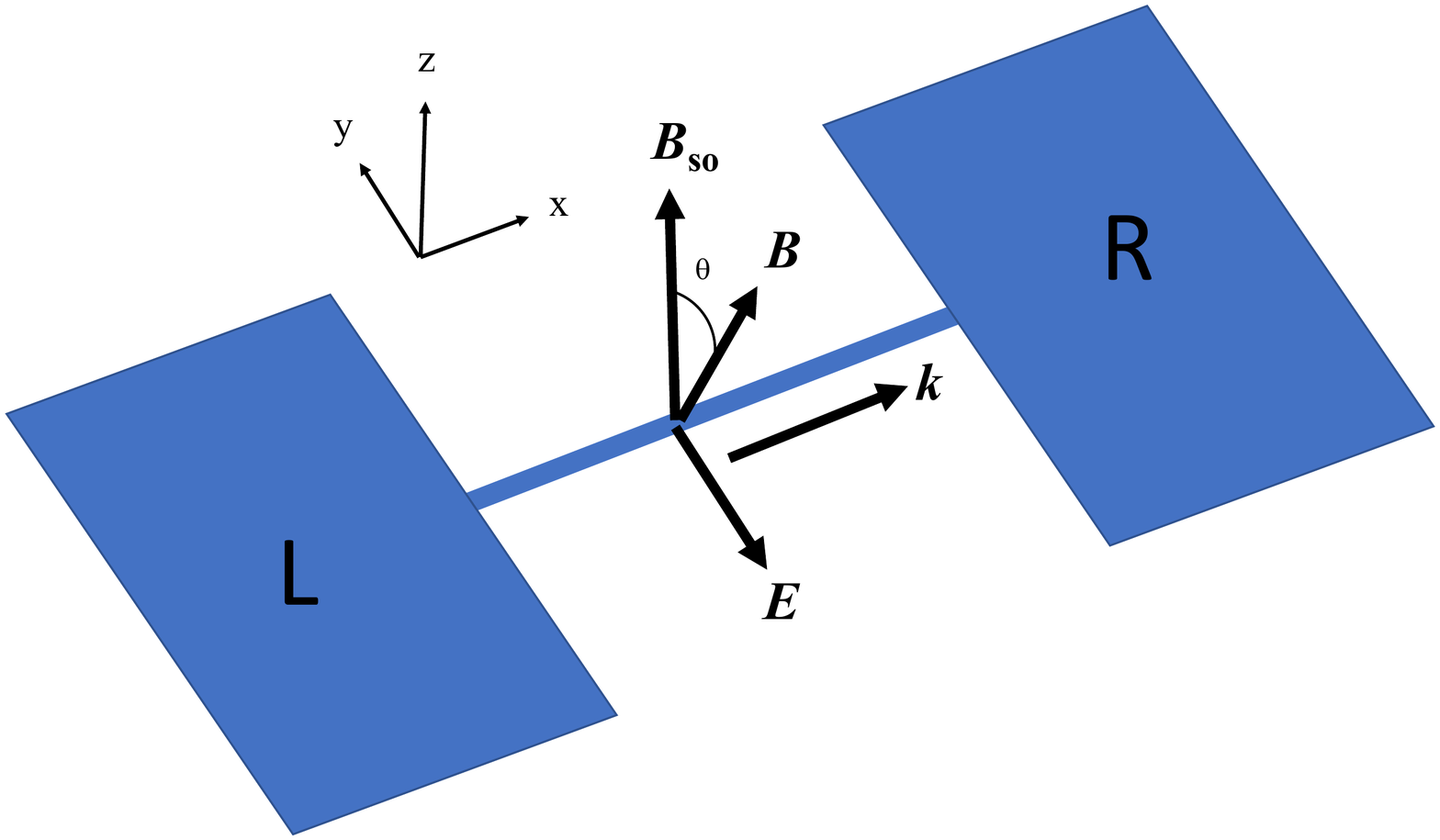}
\vspace{-1cm}
\caption{(color online) Sketch of the system  considered. A spin-orbit active one-dimensional weak link of length $d$ is attached to two leads, $L$ and $R$.
An external electric field applied in the negative $y$-direction  interacts with an electron moving along the $x$-direction with momentum $k$ and generates
a pseudo-magnetic field $B_{\rm so}$ in the $z$-direction. An external magnetic field {\bf B} is applied in the $y-z$-plane, forming an angle $\theta$
with the $z$-direction. }
\label{Fig1mj}
\end{figure}

Originally, the  Aharonov-Casher phase \cite{AC} had been predicted for a neutral particle possessing an electric moment, which moves in a magnetic field, and therefore the earlier
demonstrations  were accomplished on neutron and atomic interferometers (see e.g.,  Ref. \onlinecite{Sangster}). More recently this phase has been shown to induce changes in the  Aharonov-Bohm oscillations observed in the magnetoconductance data taken on ring structures fabricated from HgTe/HgCdTe quantum wells, where
continuously adjustable  spin-orbit coupling is available \cite{Konig}. However, the interpretation of such experiments relies 
on model calculations, and 
the deduction of that phase is rather indirect \cite{OEW}.

In this Letter we show that  the magnetoconductance  {\it anisotropy} (with respect to the direction of the magnetic field) of a {\it single }weak link allows for a detailed calibration of the Aharonov-Casher phase itself. By a weak link we here mean a pseudo-one-dimensional conductor that connects two bulk conductors and dominates the electrical resistance of the system. The system we have in mind is sketched in Fig. \ref{Fig1mj}. Electrons can propagate either ballistically or by tunneling through the weak link. We  mainly consider the ballistic  case, which is more useful in the present context, and defer a detailed discussion of the tunneling case to the supplemental material \cite{SM}.

We focus specifically on weak links where  the Rashba interaction \cite{Rashba} is controlled by an external electric field. In the absence of a magnetic field this  interaction, which preserves time-reversal symmetry, does not affect the conductance  \cite{Comment}. This is because  the two spin projections on the direction of the pseudo magnetic field, generated by the spin-orbit interaction, are good quantum numbers and hence constants of motion of the traversing electrons. As a result,  the electron transmission amplitude is diagonal in spin space and therefore the conductance does not depend on the Aharonov-Casher phase. However,
by switching on an external magnetic field, which  affects the dynamics of the electrons through the Zeeman interaction, time-reversal symmetry is broken and the above spin projections are no longer good quantum numbers.
This makes the transmission amplitude non-diagonal and mixes  the two spin projections, which correspond to different Aharonov-Casher phase factors.
The ensuing interference reflects the relative orientation of the applied magnetic field ${\bf B}$ and the pseudo field ${\bf B}_{\rm so}$ generated by the { SOI, giving  rise to an anisotropic magnetoconductance, which depends on the  product of the strength of the spin-orbit interaction
and the length of the weak link, forming the Aharonov-Casher phase.   Hence, measurements of the
anisotropy of the magnetoconductance  singles out
the Aharonov-Casher phase.


\noindent{\bf Two-terminal quantum conductance of a weak link.}
The conductance of a  conductor  coupled to two terminals
is given by the  Landauer-B\"{u}ttiker expression,
\begin{align}
G=g_{0}{\rm Tr}\{ {\bf t}{\bf t}^{\dagger}\}\ ,
\label{con}
\end{align}
$g_{0}=e^{2}/h$ being the quantum unit of  conductance \cite{Meir}.
In general the dimension of the matrix ${\bf t}$ is twice the number of channels in the link (including the two spin projections). 
For a one-dimensional weak link
the transmission amplitude ${\bf t}$ is a (2$\times 2$) matrix in spin space
\begin{align}
{\bf t}=\Gamma^{\frac{1}{2}}_{L}\G^{}_{i} (d)\Gamma^{\frac{1}{2}}_{R}\ ,
\label{amp}
\end{align}
where $\G^{}_{i}(d)$ is the Green's function (aka propagator) of the link, of length $d$,
and $\Gamma_{L}$ ($\Gamma_{R}$)
is the coupling to the left (right)  terminal, in energy units \cite{Nikolic}
(these are scalars when the terminals are 
free electron gases).

\noindent{\bf Magnetoconductance for ballistic transport.}
The explicit expression for the propagator describing ballistic transport through the weak link
reads \cite{Shahbazyan}
\begin{align}
&\G^{}_{i}(d) =\frac{d}{2\pi}\int dk e^{ikd}
\Big [E^{}_{\rm F}+i0^{+}_{}-{\cal H}(k)\Big ]^{-1}\ ,
\label{G(d)}
\end{align}
where  ${\cal H}(k)$ is the Hamiltonian of an electron moving along the  weak link (assumed to lie on the $\hat{\bf x}-$ axis) with momentum ${\bf k}=k\hat{\bf x}$ (using $\hbar=1$ units)
\begin{align}
{\cal H}(k)=\frac{k^{2}}{2m^{\ast}}-\frac{kk^{}_{\rm so}}{m^{\ast}}\sigma^{}_{z}-{\bf B}\cdot\sig\ .
\label{ham}
\end{align}
Here we have used that for electrons $\mubold = - \mu_B \sig$, where $\mu_B$ is the Bohr magneton and $\sig$ is the vector of the Pauli matrices.
The Zeeman field ${\bf B}$ (in energy units since we let $\mu_B=1$) lies in the $y-z$ plane, i.e., ${\bf B}=B\hat{\bf n}$ and $\hat{\bf n}=\sin(\theta)\hat{\bf y}+\cos(\theta)\hat{\bf z}$,
and the pseudo magnetic field induced by the spin-orbit interaction, of strength $k_{\rm so}$ in momentum units,  is along $\hat{\bf z}$. 
Comparing Eqs. (\ref{E1}) and (\ref{ham}) we note that $k_{\rm so}$ is proportional to the total electric field felt by the electron. It is common to extract its value  from experiments, see for instance Ref. \cite{Nitta}. The energy of the propagating electron is $E_{\rm F}=k^{2}_{\rm F}/(2m^{\ast})$,  $m^{\ast}$ being the effective mass of the electron.

An important point of our discussion concerns the direction of the external magnetic field, ${\bf B}$. The Hamiltonian  (\ref{ham})  shows that the effect of the spin-orbit coupling can be viewed as that of a pseudo Zeeman field which depends on the momentum and
does not break time-reversal symmetry.  We find below that when the external Zeeman field is parallel to this pseudo field (i.e., when $\theta=0$) the magnetoconductance of the weak link is totally devoid of  the Aharonov-Casher phase. As opposed, when the Zeeman field deviates away from the $\hat{\bf z}$ direction,  the magnetoconductance becomes anisotropic, with the anisotropy mainly determined by the Aharonov-Casher phase.


The integral in Eq. (\ref{G(d)}) is carried out using the Cauchy theorem \cite{SM,Shahbazyan}. It is illuminating to examine the resulting propagator 
for zero Zeeman field, 
\begin{align}
&\G^{}_{i} (d,B=0)=\frac{-im^{\ast}d}{\sqrt{k^{2}_{\rm F}+k^{2}_{\rm so}}}e^{ [id\sqrt{k^{2}_{\rm F}+k^{2}_{\rm so}}+ik^{}_{\rm  so}d\sigma^{}_{z}]}\ .
\label{B0}
\end{align}
As seen, there are two  fingerprints of the spin-orbit coupling. One is of rather  minor importance, the wave vector of the propagating electron is modified,
$k^{}_{\rm F} \Rightarrow\sqrt{k^{2}_{\rm F}+k^{2}_{\rm so}}$. The second is the accumulation of the Aharonov-Casher phase factor \cite{AC},
$\exp(i\hat\phi_{\rm AC})$, where $\hat\phi_{\rm AC}=k^{}_{\rm so}d\sigma_{z}$ is an operator in spin space.
As explained in physical terms above, and as is obvious from Eqs. (\ref{con}), (\ref{amp}), and (\ref{B0}),
the 
phase disappears from the 
conductance in the absence of an external Zeeman field and one finds that 
\begin{align}
G(B=0)/(g^{}_{0}\Gamma_{L}^{}\Gamma^{}_{R})
=2( m^{\ast}d)^{2}/(k^{2}_{\rm F}+k^{2}_{\rm so})\ .
\label{GB0}
\end{align}
A  Zeeman field with a component perpendicular to the pseudo field induced by the spin-orbit coupling is needed for this coupling to have any  effect on the conductance.

For a magnetic field along a general direction in the $y-z$ plane,  the Cauchy integration in Eq. (\ref{G(d)}) requires some care and is carried out in detail in Ref.  \cite{SM}.
Up to second order in $B_{y}$ \cite{com1} the propagator is given
by two poles at
\begin{align}
k^{}_{\pm}=\pm k^{}_{\rm so}(1-R^{}_{i}/2)+\widetilde{q}^{}_{\pm}\ ,
\label{ki}
\end{align}
where
\begin{align}
\widetilde{q}^{2}_{\pm}=k^{}_{\rm F}+k^{2}_{\rm so}\pm 2m^{\ast}B^{}_{z}+R^{}_{i}(k^{2}_{\rm so}\pm m^{\ast}B^{}_{z})\ ,
\label{q}
\end{align}
and $R_{i}=(m^{\ast}B^{}_{y})^{2}/[(m^{\ast}B^{}_{z})^{2}-(k^{}_{\rm F}k^{}_{\rm so})^{2}]$.
For zero spin-orbit coupling the propagator, and consequently the conductance,
depends solely on $B=[B^{2}_{y}+B^{2}_{z}]^{1/2}$, and is isotropic with respect to the direction of the magnetic field.
For $k_{\rm so}\neq 0$, it becomes anisotropic. Below, the conductance is calculated up to second order in the two magnetic field components,  in particular assuming that
$B_y \ll k_{\rm F}k^{}_{\rm so}/m^\ast = v^{}_{\rm F}k^{}_{\rm so}=\Delta_{\rm so}/2$. The energy split
caused by the spin-orbit interaction, $\Delta^{}_{\rm so}$, is $\approx 4-5 $ meV \cite{Nitta,Heida}.
As the magnetic 
energy (with appropriate constants reinstated) is
$g\mu_B B_y \sim g\,{\rm 0.06} (B_y/{\rm T)}\, {\rm meV}$,  this inequality might be satisfied for  reasonable values
of the $g$-factor. (Note that  the magnetic-field strength is an externally-controlled  parameter.)

Following Eqs. (\ref{con}) and (\ref{amp}), the conductance is obtained by tracing over
 the matrix product $\G^{}_{i}(d)\G^{\dagger}_{i}(d)$.
To order $(B_{z,y}/E_F)^2$ one finds that
\begin{align}
{\rm Tr}\{\G^{}_{i}(d)\G^{\dagger}_{i}(d)\}&=\frac{2(m^{\ast}d)^{2}}{k^{2}_{\rm F}+k^{2}_{\rm so}}
\Big (1+\Big [\frac{B^{}_{z}}{E^{}_{F}}\Big ]^{2}\frac{k^{4}_{\rm F}}{(k^{2}_{\rm F}+k^{2}_{\rm so})^{2}}\nonumber\\
&\hspace{-2cm}+\Big [\frac{B^{}_{y}}{E^{}_{\rm F}}\Big ]^{2}\frac{4k^{2}_{\rm F}+3k^{2}_{\rm so}}{4(k^{2}_{\rm F}+k^{2}_{\rm so})}
-\frac{1}{2}
\Big [\frac{B^{}_{y}}{E^{}_{\rm F}}\Big ]^{2}\sin^{2}(\phi_{\rm AC})
\Big )\ ,
\label{tri}
\end{align}
where the Aharonov-Casher phase, equal to $\phi_{\rm AC}=k^{}_{\rm so}d$, 
 is the eigenvalue of $\hat\phi_{\rm AC}$.
The normalized magnetoconductance is (assuming $k_{\rm so}\ll k^{}_{\rm F}$)
\begin{align}
&\frac{G-G(B=0)}{G(B=0)}
\approx \Big [1-\frac{2k^{2}_{\rm so}}{k^{2}_{\rm F}}\Big ]\Big [\frac{B}{E^{}_{F}}\Big ]^{2}\nonumber\\
&+\frac{7k^{2}_{\rm so}}{4k^{2}_{\rm F}}\Big [\frac{B^{}_{y}}{E^{}_{F}}\Big ]^{2}-\frac{1}{2}
\Big [\frac{B^{}_{y}}{E^{}_{\rm F}}\Big ]^{2}\sin^{2}(\phi_{\rm AC})\ .
\label{resi}
\end{align}
Once the Rashba interaction is switched-on, as can be done by applying gate voltages \cite{Nitta} or external electric fields \cite{Nowack}, the magnetoconductance becomes {\it anisotropic}, as demonstrated by  the last two terms in Eq.
(\ref{resi}). These  arise from different sources.
The first one is determined by the strength of the spin-orbit coupling in the particular  material forming
 the weak link. It  is  governed by the ratio $(k^{}_{\rm so}/k^{}_{\rm F})^{2}$, which is typically small: For  the values of $k_{\rm F}=\sqrt{2\pi n^{}_{s}}$ reported in Refs. \cite{Nitta,Heida} ($n_{s}$ is the electron density)
$k_{\rm F}\approx (3-6) \times 10^{8}$ m$^{-1}$ while $k_{\rm so}$ is about two orders of magnitude smaller. Thus this material-dependent anisotropy is minute
and will be neglected in what follows.

In contrast, the last term on the right hand-side of Eq. (\ref{resi}) provides a neat possibility to measure the Aharonov-Casher phase $\phi_{\rm AC}$ directly from the anisotropy of the magnetoconductance.
To do so we normalize the measured magnetoconductance for an arbitrary angle $\theta$ between the external magnetic field
[$B_y= B\sin(\theta)$, $B_z= B\cos(\theta)$]
and the pseudo field induced by the spin-orbit interaction to the measured magnetoconductance for $\theta = 0$. One finds that
\begin{align}
&\frac{G(B,\theta)-G(0,0)}{G(B, 0)-G(0,0)} = 1-\frac{1}{2}\sin^{2}(\theta)\,\sin^{2}(\phi_{\rm AC})\ ,
\label{resi1}
\end{align}
which is a result that only depends on the Aharonov-Casher phase $\phi_{\rm AC}$ and the angle $\theta$.
This function is shown in Fig. \ref{Fig1}.

\begin{figure}
\centering
\vspace{-1cm}
\includegraphics[width=.6\textwidth]{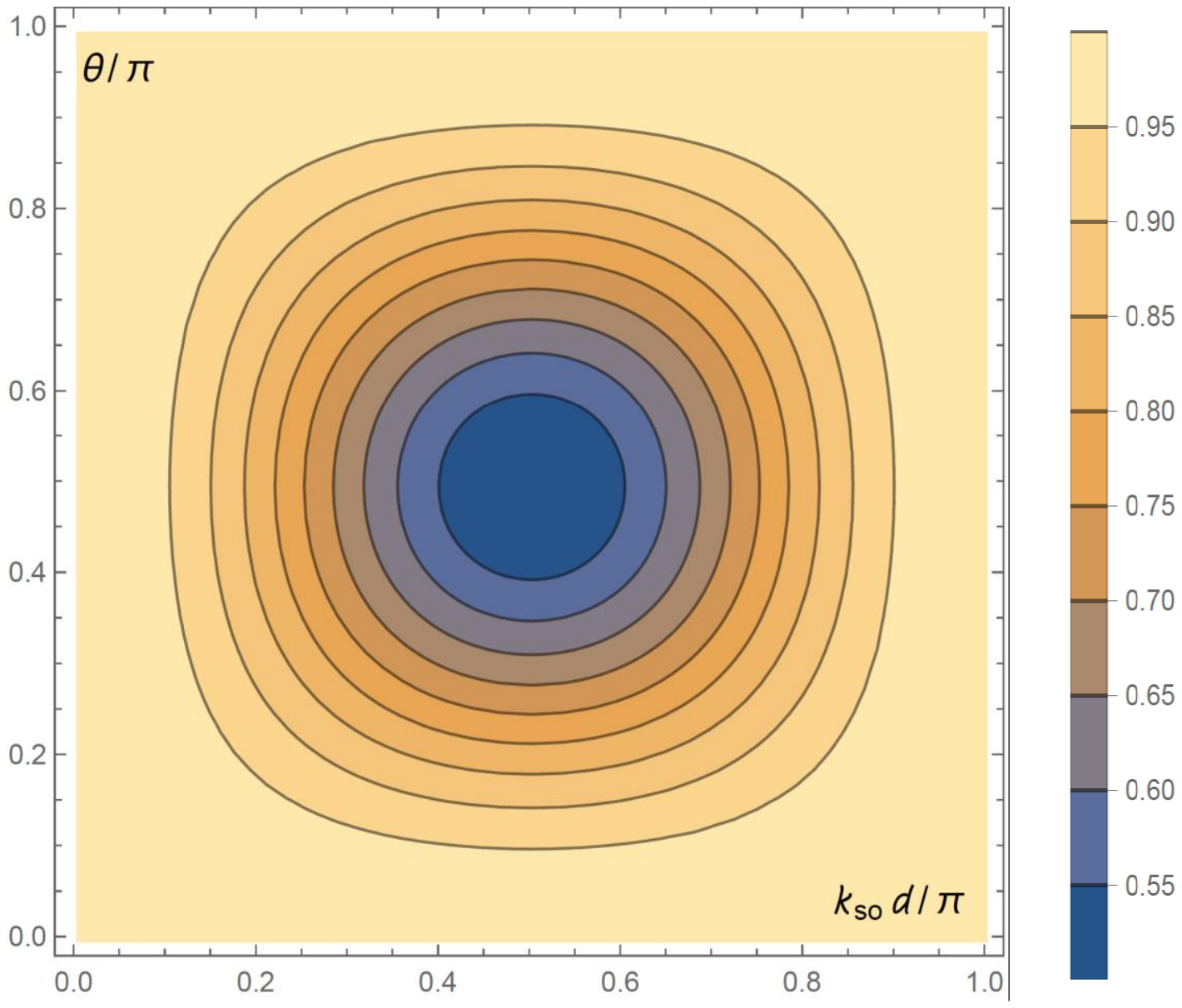}
\vspace{-8cm}
\caption{(color online) A contour plot of the anisotropy of the magnetoconductance for the ballistic case, as defined by Eq. (\ref{resi1}), plotted as a function of the
Aharonov-Casher phase $\phi_{\rm AC}=k^{}_{\rm so}d$ (horizontal axis) and the angle $\theta$ between an external magnetic field and the
pseudo-magnetic field generated by the spin-orbit interaction (vertical axis). The anisotropy has a maximum in the center of the contour plot, where $\phi_{\rm AC} = \theta =\pi/2$, and vanishes on the borders,
where $\phi_{\rm AC}$ or $\theta$ is either 0 or $\pi$. By measuring the anisotropy at different angles $\theta$ one may determine $\phi_{\rm AC}$.}
\label{Fig1}
\end{figure}


\noindent{\bf Magnetoconductance for tunneling electrons}.
The discussion above pertains to weak links through which  electrons propagate ballistically. One may wonder what form  the magnetoconductance takes for a weak link through which electron transport is by tunneling. Interestingly, the end result for the magnetoconductance anisotropy in the two cases is formally  not much different. The conductances, however, are quite disparate, implying detrimental experimental consequences.  Here we offer a heuristic comparison and explanations of the two scenarios.


Viewing a tunnel junction as a potential barrier whose height exceeds the energy of the impinging electrons by $E_0 = 1/(2m^*a_0^2)$, where 
$a_0$ can be thought of as a tunneling length, the expression for the propagator is modified. We refer to Ref. \onlinecite{SM} for a detailed derivation of the propagator $\G^{}_{e}(d)$ and for the form of ${\rm Tr}\{\G^{}_{e}(d)\G^{\dagger}_{e}(d)\}$ for this case. To order $(B_{y,z}/E_0)^2$ and for $B_y/E_0 \ll k_{\rm so}a_0 \ll 1$ one finds
\begin{align}
&{\rm Tr}\{\G^{}_{e}\G^{\dagger}_{e}\}=2(m^{\ast}a^{}_{0}d)^{2}e^{-2 d/a^{}_{0}}\nonumber\\
&\times\Big \{1+\frac{(B d)^{2}}{2(E^{}_{0}a^{}_{0})^{2}}
\Big [1+\frac{5}{2}\frac{a^{}_{0}}{d}+2\frac{a^{2}_{0}}{d^{2}_{}}\Big ]\nonumber\\
&+
\frac{(B^{}_{y}d)^{2}}{2(E^{}_{0}a^{}_{0})^{2}}\Big [\frac{a^{}_{0}}{d}\Big (\frac{\sin(\phi_{\rm AC})}{\phi_{\rm AC}} - 2\Big )
+\frac{\sin^{2}(\phi_{\rm AC})}{(\phi_{\rm AC})^{2}}- 1\Big ]\Big \}\ .
\label{Y}
\end{align}

The magnetoconductance is again anisotropic due to the term on the second line (which as expected vanishes if $B_y=0$).
The magnetoconductance, normalized as in Eq. (\ref{resi1}), becomes,
\begin{align}
&\frac{G(B,\theta)-G(0,0)}{G(B, 0)-G(0,0)} = 1 -
\frac{\sin^2(\theta)}{1 + (5/2)(a_0/d) + 2(a^{ }_{0}/d)^{2}
}\nonumber\\
&\times\Big\{
\Big [1- \Big(\frac{\sin(\phi_{\rm AC})}{\phi_{\rm AC}}\Big)^2 \Big]
+ 2\frac{a_0}{d}\Big[ 1-\frac{ \sin(2\phi_{\rm AC}) }{ 2\phi_{\rm AC}}\Big]
\Big\} \ .
\label{13}
\end{align}
In order for the anisotropy to be determined by the single parameter $\phi_{\rm AC}=k_{\rm so}d$, the Aharonov-Casher phase, as in the ballistic case,  we need 
$a_0/d \ll 1$, in which case the magnetoconductance, normalized as in Eq. (\ref{resi1}), is
\begin{align}
&\frac{G(B,\theta)-G(0,0)}{G(B, 0)-G(0,0)} 
= 1- \sin^2(\theta)\Big [1- \frac{\sin^{2}_{}(\phi_{\rm AC})}{(\phi_{\rm AC})^{2}} \Big]\ .
\end{align}
However, the assumption that $a_0/d \ll 1$ is unrealistic since that would make the conductance, which Eq. (\ref{Y}) shows is proportional to $\exp(-2d/a_0)$,
too small to be measured.
A realistic smallest value of $\exp(-2d/a_0)$ would be about 10$^{-4}$, corresponding to $a_0/d \sim 0.2$, which is a number that does not justify neglecting
terms proportional to $a_0/d$ and $(a^{}_{0}/d)^{2}$ in Eq. (\ref{13}).


Comparing the expressions for the  magnetoconductance for ballistic- and tunneling transport through a weak link,  Eqs. (\ref{tri}) and (\ref{Y}),
one observes that the
renormalization of the electronic propagator by an external magnetic field is qualitatively different
for the two transport regimes.
In the ballistic transport regime the magnetic field modifies the Fermi momentum and consequently the orbital phase factors, turning $\exp[ik^{}_{\rm F}d]$ into
$\exp[ik^{}_{\pm}d]$ [see Eqs. (\ref{B0}), (\ref{ki}),  and (\ref{q})].
These orbital phase factors have no effect on the conductance.
In a tunneling device, on the other hand, the magnetic field normalizes the tunneling length, introducing additional exponential dependence on the width $d$ of the tunnel junction. Such a modification affects the
modulus of the propagator, and reduces further the tunneling conductance.


\noindent{\bf Discussion.}
We have presented an analysis of the magnetoconductance of one-dimensional weak links through which  electrons propagate either ballistically (ballistic transport regime) or by tunneling (tunneling transport regime). 
Whether the eigenstates of the electrons in the weak link are  plane waves or  evanescent tunneling modes, 
they acquire a phase factor due to the Aharonov-Casher effect \cite{AC}.
The propagator matrix,  written in terms of these eigenstates, is  diagonal as long as the external magnetic field is parallel to the pseudo magnetic field induced by the spin-orbit coupling.
In this case the
conductance, given by Eqs. (\ref{con}) and (\ref{amp}),   is determined by the sum of the squared moduli of the diagonal elements of the propagator and therefore becomes independent of the Aharonov-Casher phase.

An external magnetic field that has a component perpendicular to the pseudo field, on the other hand, has the effect that spin is no longer a good quantum number, the propagator matrix becomes non-diagonal, and interference between its non-diagonal elements leads to a magnetoconductance that depends on the Aharonov-Casher phase. One can think of this as a manifestation of the Aharonov-Casher phase in the magnetoconductance due to an internal interference between spin-up and spin-down states. Our main result is Eq. (\ref{resi1}), showing that in the ballistic transport regime this Aharonov-Casher phase can be deduced by comparing measurements of the magnetoconductance of the weak link in an external magnetic field oriented in different directions.

\begin{acknowledgments}
We acknowledge the hospitality of the PCS at IBS, Daejeon,  Korea,
where part of this
work was done with support from  IBS  Funding No.
IBS-R024-D1. \end{acknowledgments}




\begin{thebibliography}{99}
\bibitem{com1}
The denominator in Eq. (\ref{G(d)}) is in general a quartic polynomial  with four roots, of which two contribute to the Cauchy integration. These roots can be found analytically quite simply up to order $B_{y}^{2}$ (and to all orders in $B_{z}$); see Ref. \cite{SM} for details.
\end{thebibliography}
\end{document}